\documentclass[useAMS,usenatbib]{mn2e}
\usepackage{amsfonts,amssymb}
\usepackage{graphicx}
\usepackage{ulem}
\begin{document}

\title[]{A New Dust Budget In The Large Magellanic Cloud }
\author[ Zhu et al. ]{Chunhua Zhu$^{1}$, Guoliang L\"{u}$^{1}$\thanks{E-mail:
guolianglv@xao.ac.cn (LGL)}, Zhaojun Wang$^{1}$\\
$^1$School of Physics, Xinjiang University, Urumqi, 830046,
China\\}

\date{\today}




\pagerange{\pageref{firstpage}--\pageref{lastpage}} \pubyear{}

\maketitle

\label{firstpage}

\begin{abstract}
The origin of dust in a galaxy is poorly understood. Recently, the surveys of the Large Magellanic Cloud (LMC)
provide astrophysical laboratories for the dust studies. By a method of population synthesis, we investigate
the contributions of dust produced by asymptotic giant branch (AGB) stars, common envelope (CE) ejecta and type II supernovae (SNe II)
to the total dust budget in the LMC.
Based on our models, the dust production rates (DPRs) of AGB stars in the LMC are  between about $2.5\times10^{-5}$
and $4.0\times10^{-6}M_\odot{\rm yr^{-1}}$. The uncertainty mainly results from different models for the dust yields of AGB stars.
The DPRs of CE ejecta are about $6.3\times10^{-6}$(The initial
binary fraction is 50\%).
These results are within the large scatter of several observational estimates. AGB stars mainly produce carbon grains, which
is consistent with the observations. Most of dust grains manufactured by CE ejecta are silicate and iron grains.
The contributions of SNe II are
very uncertain. Compared with SNe II without reverse shock, the DPRs of AGB stars and CE ejecta are negligible.
However, if only 2 \% of dust grains produced by SNe II can survive after reverse shock, the contributions of SNe II
are very small. The total dust masses produced by AGB stars in the LMC are between $2.8\times10^4$ and $3.2\times10^5M_\odot$,
and those produced by CE ejecta are about $6.3\times10^4$. They are much
lower than the values estimated by observations. Therefore, there should be other dust sources in the LMC.

\end{abstract}
\begin{keywords} binaries: close ¡ª stars: evolution ¡ª circumstellar matter ¡ªdust
\end{keywords}
\maketitle
\section{Introduction}
The interstellar medium (ISM) of a galaxy, one of the drivers of itself evolution,
determines many characteristics of the stars forming in it.
Dust is one of the important ingredients of the ISM, and
plays a central role in the astrophysics of the ISM. However,
the dust origin in the ISM is less understood.
According to a popular point of view,
the stellar winds of asymptotic giant branch
(AGB) stars and type II supernovae (SNe II) ejecta have long been considered
the main sites of dust formation \citep[e.g.,][]{Gail2009,Dunne2003,Gall2011}.
In addition, red supergiants, luminous blue variables, Wolf-Rayet stars and type Ia supernovae  can
also produce dust \citep[e.g.,][]{Massey2005,McDonald2009,McDonald2011,Borkowski2006,Nozawa2011}. However, compared with AGB stars and SNe II,
their contributions to the dust production are negligible \citep[e.g.,][]{Zhukovska2008,Gall2011}

Recently, as an ideal laboratory for dust formation and evolution, the Large Magellanic Cloud (LMC) was
observed by a number of surveys which detected the IR dust emissions from stellar sources
\citep[e.g.,][]{Cioni2000,Egan2001,Meixner2006,Blum2006,Ita2008,Shimonishi2013}.
Based on the observations from the {\it Spitzer Space Telescope}, \cite{Matsuura2009}
carried out the measurement of the dust production rate (DPR) by stars for the entire LMC.
They estimated that the DPR of carbon-rich (C-rich) AGB stars is $\sim$ $4.3\times10^{-5}M_\odot$ yr$^{-1}$,
much higher than that of oxygen-rich (O-rich) AGB stars (up to $\sim$ $1.7\times10^{-5}M_\odot$ yr$^{-1}$).
Based on the best fit Grid of Red Supergiant and Asymptotic Giant Branch Models, \cite{Riebel2012}
estimated that DPR by C-rich AGB stars is $\sim$ $1.3$---$1.4\times10^{-5}M_\odot$ yr$^{-1}$, which
is two and a half times as much dust into the ISM as do O-rich AGB stars.  Very recently,
based on the evolutionary models of stars through the AGB phase, \cite{Dell2014,Dell2015}
interpreted the {\it Spitzer} observations of extreme stars and AGB stars in the LMC.
Theoretically, according to the model for  dust-mass
returned from AGB stars, \cite{Zhukovska2013} calculated the DPR
of AGB stars in the LMC. They found that DPR by C-rich AGB stars is $\sim 5.7\times10^{-5}M_\odot$ yr$^{-1}$,
and DPR by O-rich AGB stars is only $\sim 1.3\times10^{-6}M_\odot$ yr$^{-1}$. Similarly, using the grids of dust yields
for different stellar masses and metallicities calculated by \cite{Ventura2012a}, \cite{Ventura2011}, \cite{Criscienzo2013}, and \cite{Ventura2014},
\cite{Schneider2014} found that DPR by C-rich AGB stars is
$\sim 4.0\times10^{-5}M_\odot$ yr$^{-1}$ ($\sim 3.0\times10^{-6}M_\odot$ yr$^{-1}$)
and DPR by O-rich AGB stars is $\sim 6.0\times10^{-6}M_\odot$ yr$^{-1}$ ($\sim 6.0\times10^{-6}M_\odot$ yr$^{-1}$), respectively.

The chemical composition of the dust mixture from AGB stars is determined
by the carbon-to-oxygen element abundance ratio ($C/O$) in the stellar wind. In O-rich environment ($C/O<1$),
dust produced is mainly silicate, while amorphous carbon dust is dominantly formed in C-rich environment ($C/O>1$).
Based on the above observational and theoretical results, dust in the LMC should be mainly composed of
amorphous carbon grains, and the silicate grains
account for a small part. However, \cite{Weingartner2001} estimated the total
volume per H atom in the carbonaceous and silicate grain populations in the Milk Way, LMC and SMC.
The volumes per H atom of the carbonaceous and silicate grain in the LMC are $\sim 0.4\times
10^{-27}$ cm$^{3}$H$^{-1}$ and $\sim 1.3\times 10^{-27}$ cm$^{3}$H$^{-1}$, respectively. Considering
the ideal graphite density of 2.24 g cm$^{-3}$  and the silicate density of 3.5 g cm$^{-3}$, we
estimate that the mass ratio of carbon to silicate grains in the LMC is about 1:4.

Therefore, most of silicate grains may not be produced by AGB stars.
Are they produced by SNe II ejecta? Unfortunately, it is presently not definitely known how much
dust is produced by SNe II ejecta. \cite{Zhukovska2008} simulated the dust returned by different sources in the solar
neighbourhood. They found that the dust grains produced by SN ejecta mainly are carbonaceous grains, while
the number of silicate grains is negligible. It is possible that most of silicate grains do not originate from SN ejecta.
\cite{Draine1979} put forward
that dust can grow in
the ISM \cite[also see][]{Draine2009}.
Unfortunately, the destruction and growth processes of dust in the ISM are poorly understood, and it is
very difficult to observe them \citep[e.g.,][]{Dwek2007}.

Recently, \cite{Lu2013}  suggested
that the common envelope(CE) ejecta in
close binary systems can provide a favourable environment for dust formation and growth.
Very recently, \cite{Nicholls2013} showed that V1309 Sco had become dominated by mid-IR emission since eruption,
which indicated the presence of a significant amount of dust in the circumstellar environment. \cite{Zhu2013}
considered that these dust grains around V1309 Sco  were efficiently produced in its progenitor-binary-merger ejecta.
The discovery of these dust grains offers an indirect evidence for the dust formation and growth in the CE ejecta.
Therefore, CE ejecta is an efficient dust source. However, to our knowledge, there is no work considering CE ejecta
as dust source on estimating dust budget in the LMC.

In this work,
we investigate the contributions of dust formed by CE ejecta to the interstellar dust,
and give a new dust budget in the LMC.
In \S 2 the model of the ejecta during binary merger is described. The main results and
discussions are presented in \S  3.  \S  4
gives conclusions.

\section{Model}
\label{sec:model}
Before discussing dust formation and growth in CE ejecta, we give some descriptions for
CE and CE ejecta evolution, and the chemical abundances of CE ejecta.

\subsection{Common Envelope Evolution}
\label{sec:ce}
As a result of dynamical timescale mass exchange in close binaries, CE
evolution plays an essential role in binary evolutions  \citep[see, e.g.,][]{Paczynski1976,Iben1993}.
Unfortunately, based on observation data and theoretical models, CE
evolution is poorly known.
In this work, we adopt the scheme by \cite{Hurley2002} to simulate CE evolution. All details can be seen in Section
2 of \cite{Hurley2002}. In most cases,
CE evolution involves a giant star (donor) transferring matter to a normal or
a degenerate star (gainer) on a dynamical timescale. If the orbital energy of the binary system is large enough,
the whole envelope of the gainer is ejected as CE ejecta.
\subsection{The Evolutions of Gas Density and Temperature in
CE Ejecta}
When CE ejecta begins to expand, its temperature and mass density start to decrease.
According to the classical
theory of nucleation \citep{Feder1966}, in most cases the formation of dust
grains can occur in adiabatically expanding gas, and is mainly determined by temperature, mass
density and chemistry of the gas phase.
In order to simulate the dust formation and growth in CE ejecta,
we must determine the evolutions of gas density and temperature.
To our knowledge,
there is no any comprehensive theoretical model or observational datum to
describe the mass density and the temperature of CE ejecta.

Considering the whole envelope is ejected
on a dynamical timescale during CE evolution, we assume that initial temperature ($T_0$) and density ($\rho_0$) of CE ejecta
are equal to the temperature and the mass density of
stellar matter in the giant envelope, respectively. In this work, the stellar structure and evolution are simulated via  a stellar evolution code developed by \cite{Eggleton1971,Eggleton1972} and \cite{Eggleton1973}, which has been updated
with the latest input physics over the past three decades \citep[e. g.,][]{Han1994}.
 The giant envelope refers to the region
 in which the hydrogen abundance (by mass) is larger than 0.5. When CE evolution occurs,
the giant envelope is ejected, shell by shell. Each shell expands outwards with no overlapping between shells.
The $T_0$ and $\rho_0$ of each shell are given by the stellar evolution code, and their evolutions are
described in the following two paragraphs.

We assume
that the CE ejecta is an adiabatically expanding perfect gas. The mass density, $\rho$, is given by
\begin{equation}
\rho={\dot{M}_{\rm ej}}/({4\pi R^2 V}),
\end{equation}
where $\dot{M}_{\rm ej}$, $R$ and $V$ are the mass-ejection rate, the radial
distance of the ejected matter and the velocity
of the ejected matter. Some numerical simulations for CE evolution give
$V\sim 100$ km s$^{-1}$, which is close to the local escape velocity\citep[e. g.,][]{Kashi2011}.
In this work, we assume that
\begin{equation}
V\simeq V_{\rm esc}=\sqrt{\frac{2GM}{R}},
\end{equation}
where $G$ is the gravitational constant and $M$ is the stellar mass.
However, considering the conservation of mass during CE expansion, we obtain
\begin{equation}
 \dot{M}_{\rm ej}=\rho4\pi R^2 V=\rho_04\pi R_0^2 V_0,
 \end{equation}
where $R_0$ is the initial distance
of a given shell of the envelope from the stellar center.
Here, $V_0=\sqrt{\frac{2GM}{R_0}}$. The gas density of the same shell at a distance $R$
from the donor center is given by
\begin{equation}
\rho=\rho_0(R/R_0)^{-3/2}.
\label{eq:rho}
\end{equation}

For an adiabatically expanding perfect gas, $\rho T^{\frac{1}{1-\gamma_{\rm ad}}}={\rm constant}$.
Based on the results of model calculations in
\cite{Fransson1989}, \cite{Kozasa1989} adopted an adiabatic index
$\gamma_{\rm ad}$ as 1.25 for the early stage of SN explosion. In this
work, we also take $\gamma_{\rm ad}$ as 1.25. Therefore, the temperature of the shell
at $R$ from the donor center is calculated by
\begin{equation}
T=T_0\left(\frac{R}{R_0}\right)^{-0.375}. \label{eq:tem}
\end{equation}
Eqs. (1)---(5) determine the evolution of mass density and temperature of every shell in CE ejecta.
\subsection{Chemical Abundances of CE Ejecta}

The chemical abundances of CE ejecta are determined by the donor's envelope.
For a single star, three dredge-up processes and hot bottom
burning (the latter only in a star with initial mass higher than 4Msun) may change the chemical abundances of the stellar
envelope. During each dredge-up event the base of the convective envelope extends inwards to shells previously involved in nuclear burning and nuclearly processed matter is transported to the external zones of the star\citep[e. g.,][]{Iben1983,Herwig2005}.
The first dredge-up occurs during the first ascent of the
giant branch (FGB). Sufficiently massive stars ($M_{\rm i}>4M_\odot$) undergo the second dredge-up
after the end of He-core burning\citep[e. g.,][]{Herwig2005}. The third dredge-up occurs
when the stars evolve to thermal pulse AGB (TPAGB) phase. If the mass of the hydrogen envelope above the
He-exhausted core is large enough and the core mass is above $\sim 0.8 M_{\odot}$, the hydrogen burning shell can extend into
the bottom of the convective region, which is called as hot bottom burning.
Reviews about dredge-ups and hot bottom burning were given by \cite{Iben1983} and \cite{Herwig2005}.

As discussed in the previous section, we use Eggleton's code to simulate the evolution of a star
from the premain sequence to the end of the AGB phase. However, this code does not simulate the
effects of third dredge-up process and hot bottom burning on the chemistry of stellar envelope.
In this work, we use the model of \cite{Lu2008} to simulate the chemical
evolutions of $^1$H, $^4$He, $^{12}$C, $^{13}$C, $^{14}$N, $^{15}$N,
$^{16}$O, $^{17}$O, $^{20}$Ne and $^{22}$Ne in the stellar envelope.
\cite{Lu2008} followed the prescriptions  by
\cite{Izzard2004} for the first dredge-up during FGB and
the second dredge-up during early AGB.
For the third dredge-up and the hot bottom burning during the
TPAGB phase, we use the TPAGB synthesis in \cite{Groenewegen1993},
\cite{Karakas2002}, \cite{Izzard2004} and \cite{Marigo2007}.
All details can be found in Appendix A of \cite{Lu2008}.

Mass transfer in a binary system can change the chemical abundances
of the stellar envelope. In binary
systems, there are two ways to transfer mass: (i)accretion from the
stellar wind material of a companion star; (ii)Roche lobe overflow.
After the gainer in a binary system obtains mass ( $\Delta M$) from the donor,
the chemical abundances  of the stellar surface, $X_{\rm g}$, are
given by
\begin{equation}
X^{\rm new}_{\rm g}=\frac{X^{\rm old}_{\rm g}\times M^{\rm
env}_{\rm g}+X_{\rm d}\times\Delta M}{M^{\rm env}_{\rm g}+\Delta M},
\end{equation}
where $M^{\rm env}$ is the envelope mass and $X$ is the chemical
abundances of its companion star, g and d indicate the gainer and the donor in a binary
system, respectively.

Then, we follow the chemical
evolutions of $^1$H, $^4$He, $^{12}$C, $^{13}$C, $^{14}$N, $^{15}$N,
$^{16}$O, $^{17}$O, $^{20}$Ne and $^{22}$Ne in the stellar envelope of binary system.
The abundance ratio of carbon to oxygen, $C/O=(^{12}C+ ^{13}C)/(^{16}O + ^{17}O)$,
determines the dust species produced by
CE ejecta.
O-rich ejecta($C/O<1$) forms silicate dust, while C-rich ejecta($C/O>1$) produce
amorphous carbon and SiC. Following \cite{Ferrarotti2006}, we assume that the abundances of other key elements (Fe, Si, Mg and S)
 for the dust formation do not change.

\begin{figure}
\includegraphics[totalheight=3.0in,width=2.5in,angle=-90]{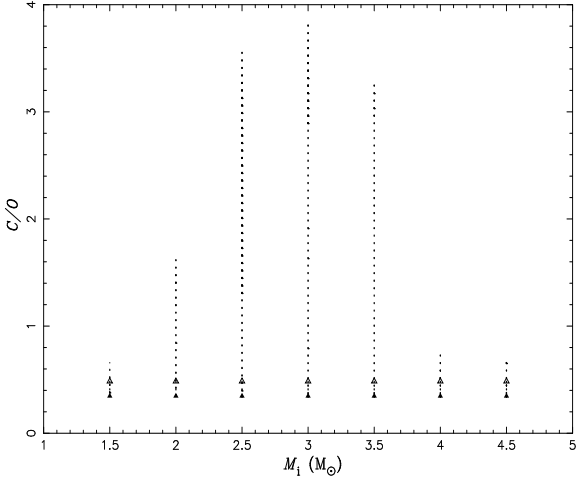}
\caption{The evolutions of $C/O$s in the stellar envelopes with
different initial masses. The dotted lines represent the evolutions of $C/O$s.
The empty and solid triangles give the initial $C/O$s and the $C/O$s after stars undergo the first dredge-up (FDU), respectively.
The tops of dotted lines
represent the maximum values of $C/O$s at the end of AGB phase.
 }
\label{fig:abun}
\end{figure}

 Following \cite{Lu2008}, we use a synthetic
stellar evolution model to calculate the evolution of chemical abundances.
Figure \ref{fig:abun} shows the time evolution of the surface C/O of stars with different masses.
The evolutions of $C/O$ mainly depend on the initial masses of the stars:\\
(1) All stars undergo the first dredge-up during the FGB phase. Its effects mainly are that carbon abundance on the stellar surface
is reduced by approximately 30\% and oxygen abundance does not change. As Figure \ref{fig:abun} shows, $C/O$s of all
stars decrease after the first dredge-up.
The second dredge-up, occurring in models with mass above $4 M_{\odot}$, hardly changes the $C/O$ ratio\citep[e. g. ,][]{Izzard2004}.\\
(2)Low-mass stars ($M<1.5M_\odot$) never become carbon stars, owing to the
small number of thermal pulses experienced. \\
(3)Stars with initial masses between about 1.5 $M_\odot$ and 4.0 $M_\odot$
increase their carbon abundance over the oxygen abundance after a number of thermal
pulses. The surface $C/O$ soon exceeds unity. \\
(4)Intermediate mass stars with initial masses between about 4 $M_\odot$ and 8 $M_\odot$
convert the dredged-up carbon rapidly into $^{14}$N via the CN-cycle due to hot bottom burning.
However, the oxygen is not affected by this process.  Therefore, the carbon abundance in the
envelopes of these stars is much less than the oxygen abundance. These stars never become carbon stars.\\

\subsection{Dust Formation and Growth in CE Ejecta}
With the decrease of CE ejecta's temperature due to its expansion, dust
can form and grow. In this work, with the help of {\it AGBDUST} code (kindly provided by Prof. Gail, private communication),
we replace the thermodynamic structure of the wind from
AGB star in the {\it AGBDUST} code with Eqs. (\ref{eq:rho}) and (\ref{eq:tem}), and
use the model of formation and
growth of dust grains in  the {\it AGBDUST} code to simulate the formation and
growth of dust grains in CE ejecta.

The model of formation and
growth of dust grains in the {\it AGBDUST} code assumes that the dust grains grow
on some kinds of seed nuclei. The seed nuclei are formed prior to the point where the different dust
components start to condense. The radii of these seed nuclei are denoted by $a_{\rm 0}$.
The {\it AGBDUST} code assumes $a_0=1$ nm for the radii of the seed nuclei. If the wind is accelerated by radiation pressure,
the final results for the grain radii are nearly independent of the particular assumption on
$a_0$.
Then, the growth of the grain size of species $i$ is calculated
via the equation
\begin{equation}
\frac{{\rm d}a_{i}}{{\rm d} t}=V_{0,i}(J^{\rm gr}_{i}-J^{\rm dec}_{i}),
\label{eq:da}
\end{equation}
where $a_{i}$ is the radius of dust species $i$,
and $V_{0,i}$ is the volume of the nominal molecule in the solid.
Here, $J^{\rm gr}_{i}$ is the deposition rate per unit time and
surface area for a particle in rest with respect to the gas phase.
It is given by
\begin{equation}
J^{\rm gr}_{i}=\alpha_i n_i v_{{\rm th},i},
\label{eq:jgr}
\end{equation}
where $\alpha_i$ denotes the sticking coefficient, $n_i$ gives the particle density in the
gas phase and $v_{{\rm th},i}=\sqrt{\frac{kT}{2m_i}}$ is the mean thermal velocity of the
particles. Here, $k$ is the Boltzmann constant and $T$ is the temperature.
The $J^{\rm dec}_{i}$ is decomposition rate
in a thermodynamic equilibrium state, which is given by
\begin{equation}
J^{\rm dec}_{i}=\alpha_i v_{{\rm th},i} \frac{p_i}{kT},
\end{equation}
where $p_i$ is the partial pressure of the molecules of dust species $i$  in chemical
equilibrium between the gas phase and the solid phase.

The amount of dust formed is indicated by the fraction
of gas condensed into solid grains, i.e. the degree of condensation\citep{Gail1999}.
For each species, the {\it AGBDUST} code
selects a key-element $k$. Its degree of condensation is expressed by, $f^i_k$ , the
fraction of the key-element in dust species $i$, which is given by
\begin{equation}
f^i_k=\frac{4\pi[(a_i)^3-(a_0)^3]}{3V_{0,i}}\frac{n_{{\rm d},i}}{\epsilon_k N_{\rm H}},
\end{equation}
where $n_{{\rm d},i}$ is the number density of seed nuclei of the different
dust species, $\epsilon_k$ is the key-element abundance by number relative to H abundance,
and $N_{\rm H}$ is the number density of hydrogen nuclei. In the {\it AGBDUST} code,
$n_{{\rm d},i}=3\times10^{-13}N_{\rm H}$ for all dust species \citep{Ferrarotti2003}.
Based on the degrees of condensation of key-element ($f^i_k$) in dust species $i$, the DPR
in CE ejecta is calculated by
\begin{equation}
\frac{{\rm d}M_{\rm d}^{i}}{{\rm d}t}=\dot{M_{\rm ej}}X_{k}\frac{A_{i}}{A_{k}}f^i_k,
\label{eq:odu}
\end{equation}
where $X_{k}$ is the key-element abundance
in CE ejecta by mass, $A_{i}$ and $A_{k}$ are the
molecular and atomic weights of the dust species $i$ and the key-element $k$, respectively.

Table \ref{tab:dust} gives a list of the dust species considered in the present paper, their formation reactions, the corresponding key-elements and
the sticking coefficient ($\alpha_i$). The thermodynamic quantities
(mainly the free energies used to calculate the equilibrium pressures) of the reactions are taken from \cite{Sharp1990}.
The composition of olivine and pyroxene depends on
the relative fraction of magnesium and iron, indicated, with $x$ and $1-x$(see Table \ref{tab:dust}), respectively.
The details for calculating the magnesium percentages in the olivine and pyroxene dust grains can be found in \cite{Gail1999}.
To our knowledge, there is not experimentally determined data for the sticking coefficient. We adopt the values
of $\alpha_i$ in \cite{Ferrarotti2003} which are mainly based on \cite{Hashimoto1990} and \cite{Nagahara1996}.

\begin{table*}
\centering
  \caption{Dust species considered in the present paper, their formation reactions, the corresponding key-elements,
  and the sticking coefficient ($\alpha_i$).}
  \tabcolsep2.50mm
 \begin{tabular*}{140mm}{cccc}
 \cline{1-4}
\hline\cline{1-4} \hline
 Dust Species & Formation Reactions & Key-Elements &$\alpha_i$\\
 Olivine &${\rm 2{\it x}Mg +2(1-{\it x})Fe+SiO+3H_2O\rightarrow Mg_{2{\it x}}Fe_{2(1-{\it x})}SiO_4 + 3H_2}$ & Si & 0.15\\
 Pyroxene&${\rm {\it x}Mg +(1-{\it x})Fe+SiO+2H_2O\rightarrow Mg_{\it x}Fe_{(1-{\it x})}SiO_3 + 2H_2}$& Si & 0.15 \\
 Quartz  &${\rm SiO + H_2O\rightarrow SiO_2(s) +H_2}$& Si & 0.1 \\
 Silicon Carbide &${\rm 2Si + C2H2 \rightarrow 2SiC + H_2 }$& Si & 0.3 \\
Carbon &${\rm C\rightarrow C(s)}$& C & 0.3 \\
Iron &${\rm Fe\rightarrow Fe(s)}$& Fe & 0.8 \\
 \cline{1-4}
 \label{tab:dust}
\end{tabular*}
\end{table*}

\section{Dust Yields from CE Ejecta}
Based on the model of dust formation and growth in CE ejecta described in the last section,
dust yields not only depend on the parameters listed in Table \ref{tab:dust},
but also on the initial temperature ($T_0$) and density ($\rho_0$),
the chemical abundances (mainly $C/O$) and the mass of CE ejecta.

In this section, we
take $Z=0.008$ as an example to calculate the dust yields from CE ejecta.
The initial abundances of
all heavy elements are approximately equal to 0.4 times
those of the Sun. The element abundances on
the surface of the Sun are taken from \cite{Anders1989}.

\subsection{Typical Examples}
As discussed in \S \ref{sec:model}, CE evolution involves a giant or giant-like envelope.
This giant star may evolve in the Hertzsprung gap (HZ), FGB or AGB.
The envelopes at different evolutionary stages
have different structures (mass density, temperature and chemical abundances),
which may affect the dust formation and growth in the CE ejecta. In order to discuss these effects,
we select two groups of typical examples: one is that CE ejecta originate from an envelope
of a star with initial mass of $1 M_\odot$
at the beginning of HZ, the beginning of FGB, the end of FGB, the beginning of AGB and TPAGB, respectively; the other is that CE
ejecta is produced by an envelope of a star with initial mass of $3 M_\odot$
at the begin of HZ, the begin of FGB, the end of FGB, the begin of AGB, and $C/O=2$ at TPAGB due to the third dredge-up, respectively.

\begin{figure*}
\begin{tabular}{cc}
\includegraphics[totalheight=2.5in,width=2.5in,angle=-90]{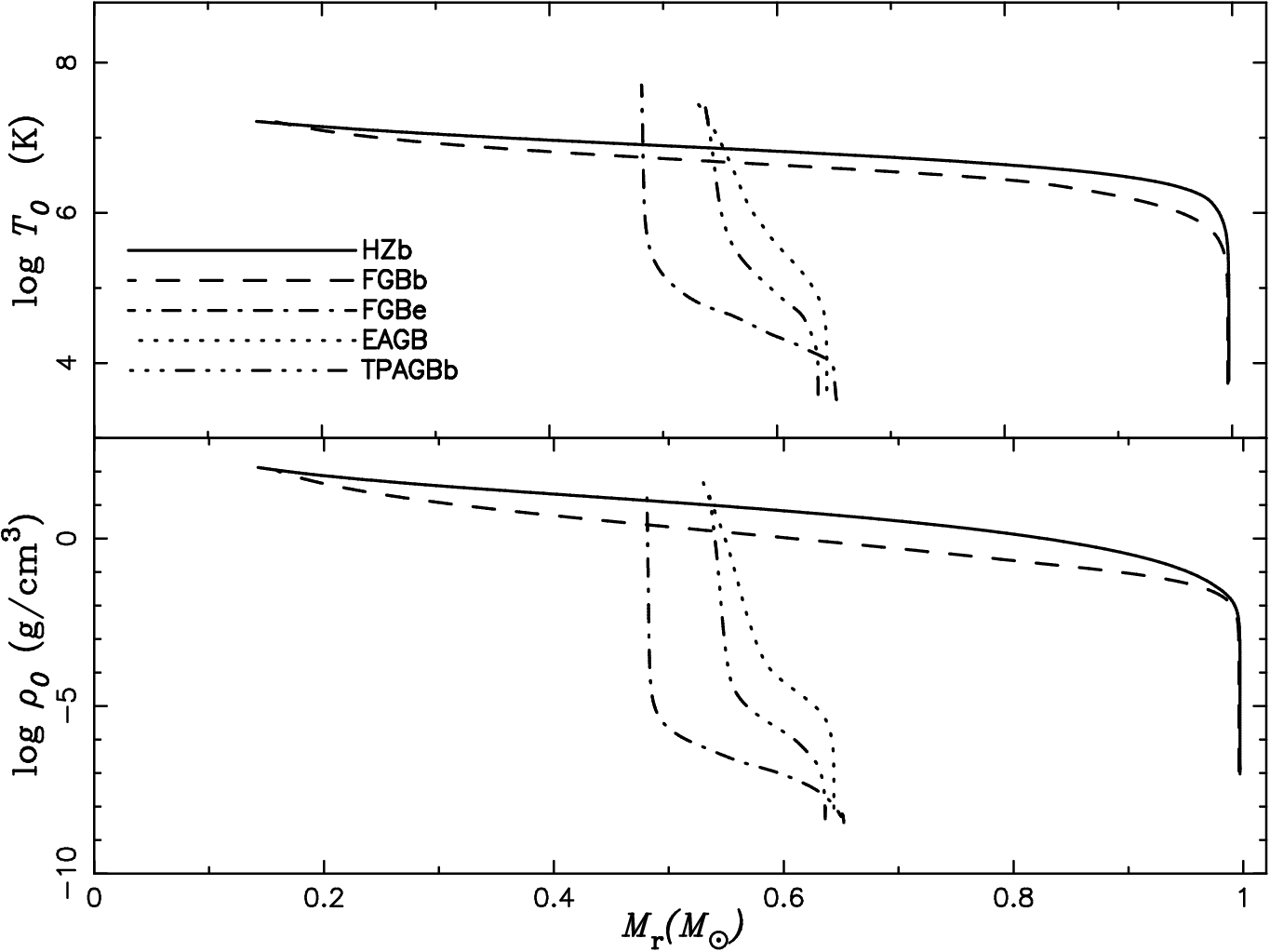}
&\includegraphics[totalheight=2.5in,width=2.5in,angle=-90]{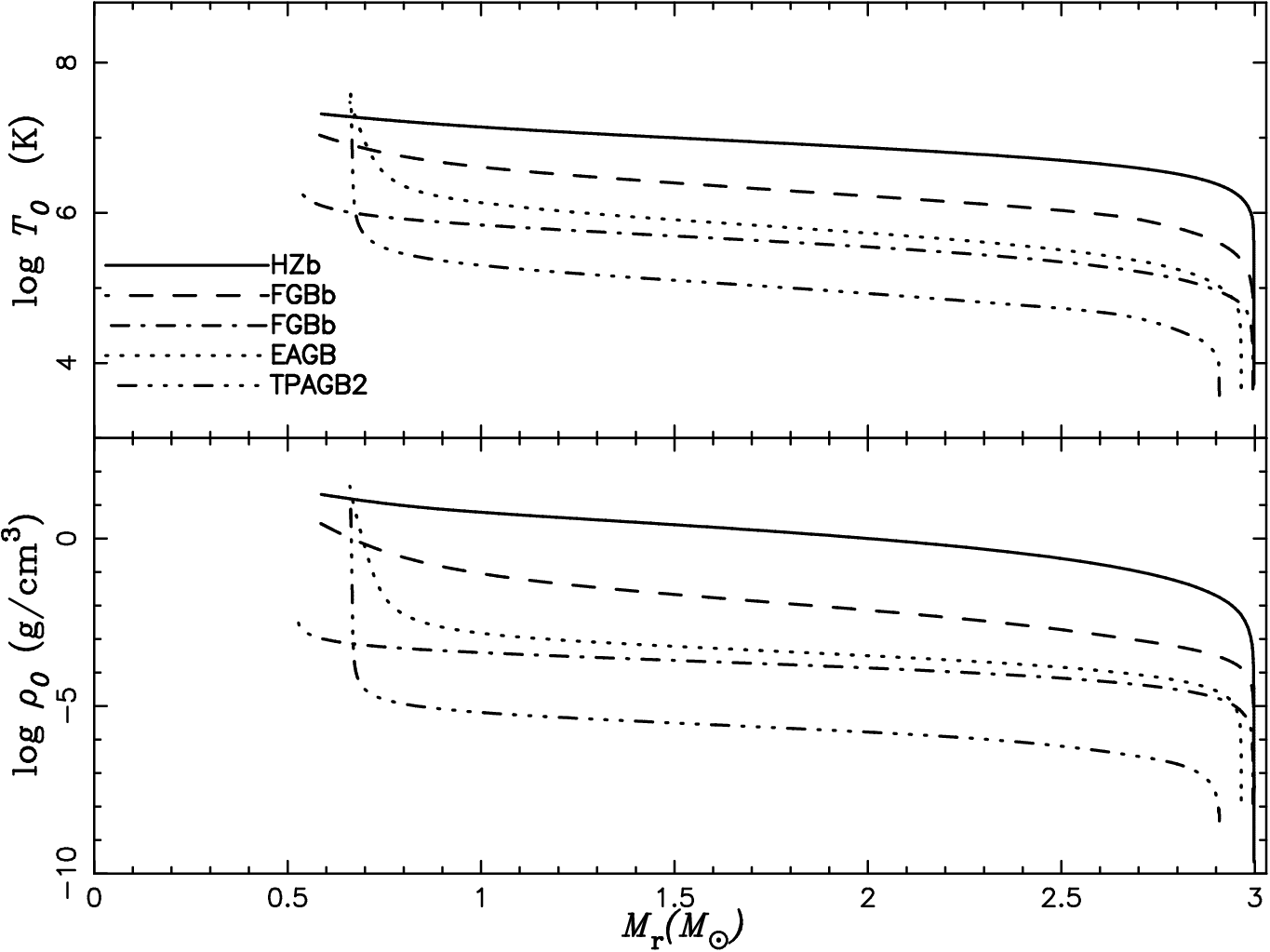}\\
\end{tabular}
\caption{The initial temperature and density profiles of the
envelopes of giants (at different evolutionary stages as indicated)
as a function of relative mass coordinate. 'HZb',  'FGBb', 'FGBe',
'EAGB' and 'TPAGBb' mean that star just evolves at the begin of Hertzsprung gap,
the begin of the FGB, the end of the FGB, the begin of the AGB and the begin of
the thermally pulsing AGB (TPAGB), respectively. 'TPAGB2' means that $C/O$ of stellar envelope increases to
2.0 because of the third dredge-up. The left and right panels are for
the star with mass of 1 and 3 $M_\odot$ at zero-age main sequence, respectively.}
\label{fig:tero}
\end{figure*}

Figure \ref{fig:tero} shows the initial temperature and density profiles of the
giant envelopes as a function of relative mass coordinate for the two groups of typical examples.
The external regions of the stars expand during the evolution, with the consequent
decrease of temperature and density. The mass of the envelope diminishes, owing to mass
loss.

\begin{figure*}
\includegraphics[totalheight=6.2in,width=3.5in,angle=-90]{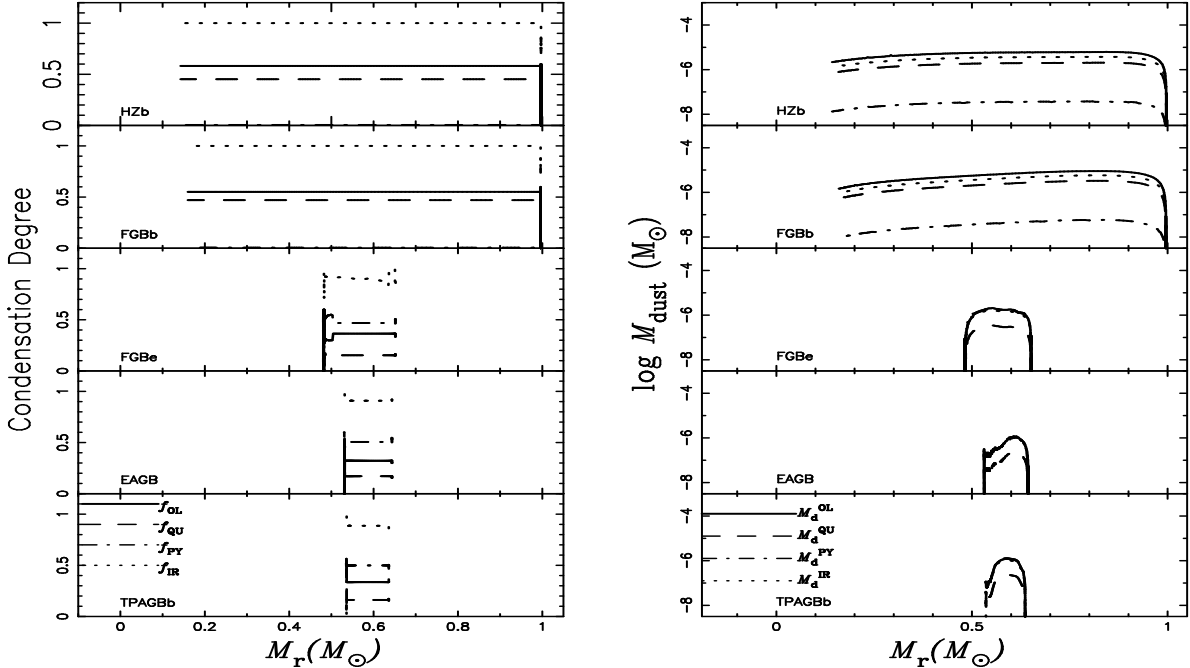}
\caption{Condensation degrees of key-elements and dust yields of different dust species
as a function of relative mass coordinate at different evolutionary stages for
the star with mass of 1 $M_\odot$
when stellar envelope is ejected into a region where dust can form and grow(Details in text).
The '$f_{\rm OL}$', '$f_{\rm QU}$' and '$f_{\rm PY}$'
present the condensation degrees of Si element in olivine-type, pyroxene-type and quartz grains, respectively.
The '$f_{\rm IR}$' means the condensation degree of Fe element in iron grains. Correspondingly,
the '$M_{\rm d}$'s present the dust yields of different dust species. }
\label{fig:cond1}
\end{figure*}

\begin{figure*}
\includegraphics[totalheight=6.2in,width=3.5in,angle=-90]{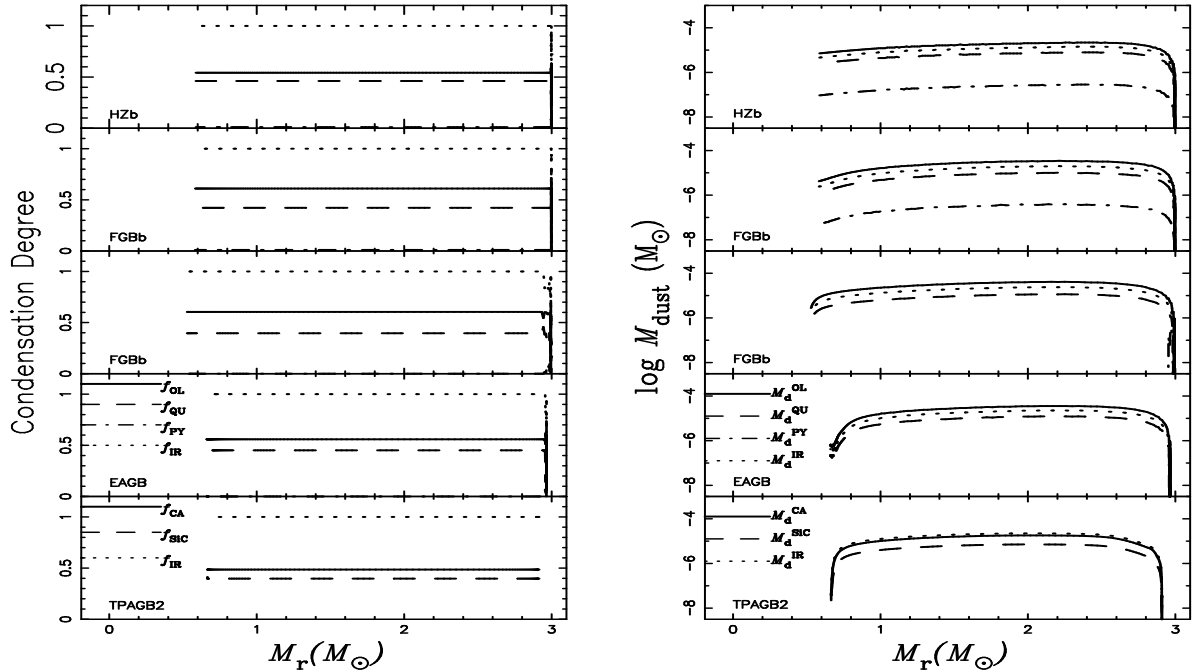}
\caption{Similar to Figure \ref{fig:cond1}, but for the star with mass of 3 $M_\odot$.
The '$f_{\rm CA}$' and '$f_{\rm SiC}$' present the condensation degrees of carbon and Si elements in
carbon and silicon carbide grains, respectively.  }
\label{fig:cond3}
\end{figure*}

As shown in Figure \ref{fig:tero}, at the beginning of CE expansion the temperatures in the envelope
are too large (higher than $\sim 10^4$ K) to allow dust formation. During the expansion the CE ejecta reach regions
$10^{13}$---$10^{16}$ cm away from the donor, where the temperatures are below 2000K, thus rendering
possible condensation of gas molecules into dust\citep{Lu2013}. By means of the
{\it AGBDUST} code we simulate dust formation and growth within the expanding envelope.
The computation is run for each shell, of initial temperature $T_0$, into which the envelope
is divided. Figures \ref{fig:cond1} and \ref{fig:cond3} show the condensation degrees of key elements and the mass of
dust formed in the expanding shells of the envelope of two stars of mass 1 and 3 solar masses.

The dust formation and growth in CE ejecta
has its own features: \\
(i)High condensation degrees. The main reason is that the dust-forming zones in the CE ejecta ($T<\sim 1300$ K) have
very high mass density ($\sim 3.3\times10^{-11}{\rm g\ cm}^{-3}$). Based on Eqs.(\ref{eq:da}) and (\ref{eq:jgr}),
 the higher is the mass density, the higher is the growth rate of grain size, which results in a high condensate degree.
For AGB
wind with a high mass-loss rate of $1\times10^{-5}\, M_\odot\, {\rm
  yr}^{-1}$ and a wind structure based on \cite{Ferrarotti2006},
{\it AGBDUST} code shows that the temperature and the mass density of dust-forming
zones are $\sim$ $1.3\times10^{3}$ K and $5.0\times10^{-14}{\rm
  g\ cm}^{-3}$, respectively. About 32\% of Si atoms in  AGB
wind condensate into silicates, and
about 4\% of Fe atoms condensate into iron grains. However, in our simulations,
all of Si and Fe atoms in CE ejcta condensate into silicate
($f_{\rm OL}+f_{\rm QU}+f_{\rm PY}\sim 1$) and iron grains ($f_{\rm IR}\sim 1$), respectively.
 \\
(ii)Silicate grains are mainly made up of pure forsterite (${\rm Mg_{2}SiO_4}$) and quartz grains.
At the region of dust formation and growth, compared with AGB stellar wind, CE ejecta has higher
temperature and mass density. Based on the stability limits of some grains and molecules
(See Fig. 2 in \cite{Gail1999}), forsterite, iron and quartz grains firstly form.
Most of Fe atoms condensate into iron grains. Therefore,
olivine-type (${\rm Mg_{2{\it x}}Fe_{2(1-{\it x})}SiO_4}$) grains are mainly made up of
pure forsterite (${\rm Mg_{2}SiO_4}$) grains.\\

As Figures \ref{fig:cond1} and \ref{fig:cond3} show, almost all of
Fe atoms in CE ejecta condensate into
iron grains. In a CE ejecta ($C/O<1$), most of Si element condensates into silicate grains;
In CE ejecta ($C/O>1$),  a part of C atoms and most of O atoms combine CO molecule,
while most of left C atoms condensate into carbon grains.  Therefore, CE ejecta
can efficiently produce dust grains, and the dust mass mainly depends on the
mass of CE ejecta.

\subsection{Comparison with AGB Stars}
AGB stars, as important stellar sources of dust, have been investigated by many independent groups.
Having used synthetic stellar evolution models, \cite{Zhukovska2008} computed dust condensation and estimated
the dust yields for M, S, and C-type AGB stars of different ages and metallicities. Based on these models,
\cite{Zhukovska2013} estimated dust input from AGB stars in the LMC. However, as \cite{Schneider2014} pointed out,
their major drawback was that they used simple evolutionary code based on analytic approximations to describe
complex physical processes, primarily third dredge-up and hot bottom burning. These processes directly affect the
chemical abundances and physical conditions in  AGB envelope, thus they determine the dust yields.
Ventura and collaborators used the ATON code to produce accurate modelling of the AGB phase
and computed the mass and composition of dust produced by AGB stars
(\cite{Ventura2011}; \cite{Ventura2012a,Criscienzo2013}). Following \cite{Schneider2014}, we call this code as old ATON.
 \cite{Zhukovska2013} found that, compared with observations,
the old ATON yields underestimated the dust production rate from C-rich AGB stars.
For this reason, \cite{Ventura2014} calculated new AGB evolutionary sequences with a deeper third dredge-up,
and \cite{Schneider2014} provided new dust yields, to which we refer to as new ATON. However,
there are some differences of dust mass and composition among \cite{Zhukovska2008}, old and new ATON.
Detailed reasons had been discussed by \cite{Schneider2014} and \cite{Ventura2014}.

As discussed in next section, based on the observational data,
several investigations estimated the DPRs from AGB stars in the LMC, and their results
have large scatter.
By comparing the observed and predicted DPRs of the Magellanic
Clouds, \cite{Schneider2014} found that old ATON models are better suited for $Z<=0.004$, while use of the new ATON
models is recommended for $Z=0.008$. In this work, in order to compare the dust yields produced by CE ejecta
with those by AGB stars, we use the above three models.

\begin{figure*}
\includegraphics[totalheight=5.50in,width=3.5in,angle=-90]{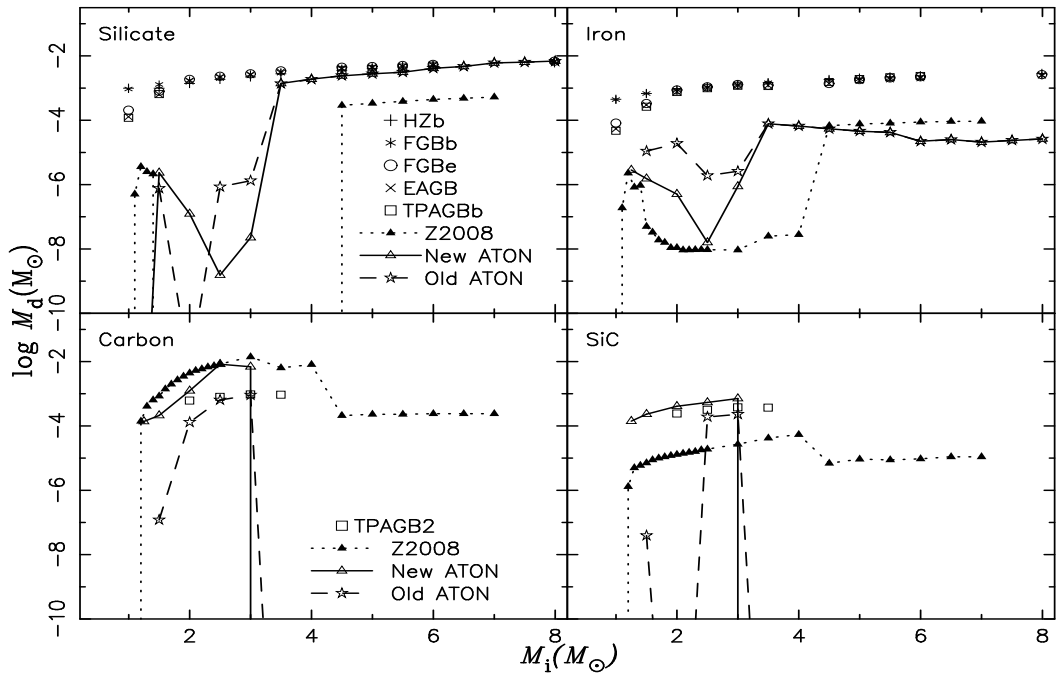}
\caption{Comparison of the dust yields from CE ejecta formed at
  different evolutionary stages with AGB stellar winds. Z2008, old and new ATON mean that
  the data come from \citet{Zhukovska2008}, old ATON (\citet{Ventura2012a}) and new ATON (\citet{Schneider2014}), respectively.}
\label{fig:qdust}
\end{figure*}

In Figure \ref{fig:qdust}, we compare the dust masses of different compositions from CE ejecta formed
at different evolutionary stages with those from AGB stars by \cite{Zhukovska2008}, old
and new ATON.
Due to the high condensation degrees in almost the whole envelope,
CE ejecta produces more efficiently silicate and iron grains.
In our model, the stars with initial masses between $\sim$ 1.5 and 4.0 $M_\odot$ evolve to carbon stars and
 produce carbon and SiC grains, similar to the results of old and new ATON. On the other hand,
having used a model for TPAGB with efficient third dredge-up and weak hot bottom burning, the stars between $\sim$ 1 and 8 $M_\odot$
in \cite{Zhukovska2008} can become carbon stars with high $C/O$ at the end of AGB stars. Therefore, their models
produced more carbon grains and less silicates.

\subsection{Comparison with Supernova}
SNe II are considered as another important stellar sources of dust.
Due to their short lifetimes and large production of metals,
they are regarded as the most likely sources of dust in high red-shift galaxies and quasars
\citep[e.g.,][]{Dwek1998,Gall2011}. Unfortunately, no matter observations or theoretical models,
there is no reliable result to definitively calculate which species of dust formed and in what quantities
by SNe II. Several theoretical models for dust formation in the ejecta of SNe II have been
developed \citep[e.g.,][]{Todini2001, Nozawa2003}. Based on standard nucleation theory, \cite{Todini2001}
calculated dust yields in the ejecta of SNe II whose progenitors have different metallicities
($Z=0, Z=10^{-4}, Z=10^{-2}$ and $Z=0.02$). By a linear interpolation between $Z=10^{-4}$ and $Z=10^{-2}$,
we can obtain dust yields from SNe II whose progenitors have $Z=0.008$.
Every SN can produce a mass of dust of up to 1 $M_\odot$. However, \cite{Bianchi2007} reconsidered the model of \cite{Todini2001} and
included dust destruction. They found that only about 2 --- 20 percent of the initial
dust mass can survive the passage of the reverse shock, depending on
the density of the surrounding interstellar medium.

\begin{figure}
\includegraphics[totalheight=3.0in,width=2.5in,angle=-90]{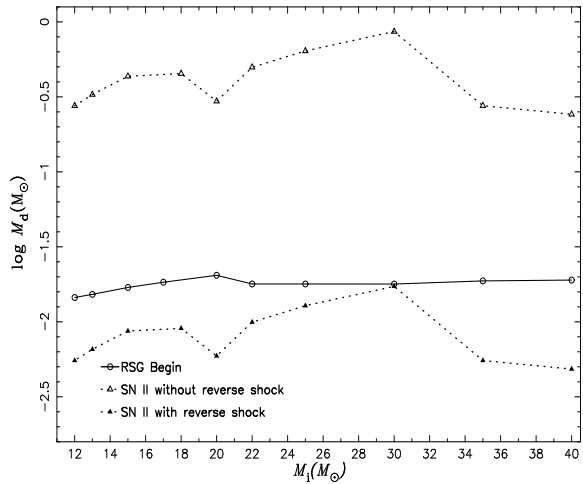}
\caption{Comparison of the dust yields from CE ejecta formed by massive stars at the begin of red supergiant (RSG).
The dust yields from SNe II without reverse shock come from \citet{Todini2001}.
The dust yields from SNe II with reverse shock mean that only about 2 percent of dust mass survive after reverse shock.
 }
\label{fig:mdust}
\end{figure}

In a binary system, if a massive star fills its Roche lobe when it evolves into the red supergiant (RSG) state,
its envelope can also be ejected as CE ejecta. We compare in Figure \ref{fig:mdust} the dust yields in CE ejecta with those from SNe II.
In the ejecta of SNe II, there is a large quantity of metal elements.
In the model of \cite{Todini2001}, most of them condensate into dust grains. Therefore, compared with
the ejecta of SNe II without dust destruction, dust produced by CE ejecta from massive star is negligible.
However, as \cite{Bianchi2007} showed, most of dust produced by the ejecta of SNe II are soon destroyed by
reverse shock. If only about 2 percent of the initial
dust mass produced by the ejecta of SNe II can survive after the reverse shock, dust produced by CE ejecta from massive star is dominant.


\section{Dust Budget in the LMC}
As shown in the last section, AGB stars, the ejecta from SNe II and CE are
potential sources of dust in the LMC. In order to calculate their contributions
to total dust in the LMC, we use a method of population synthesis.

\subsection{Basic Parameters of Population Synthesis}
\label{sec:monte}
For a method of population synthesis, the main input
parameters are:(i)the initial mass function (IMF);
(ii)the mass-ratio distribution for binaries;(iii)the eccentricity distribution for binaries;(iv)the distribution
of orbital separations for binaries. We use a Salpeter initial mass-function for single stars and primary components in binaries:
\begin{equation}
\phi(m)=C m^{-2.35},
\end{equation}
with normalisation number $C=0.06$ within the mass integral range between 0.1 and 100 $M_\odot$.
For binary systems, we must know the distribution of the ratios of components.
Based on \cite{1989Ap.....30..323K} and \cite{1994A&A...282..801G},
a flat distribution of initial mass ratios of components is assumed in this work.
Meanwhile we assume that all binaries have initially circular orbits.
In our work, one of the most important input parameters is the distribution of initial binary separations
which directly determine how many binaries undergo Roche-lobe overflow. However, to our knowledge, there
is no any confirmed observational evidence to describe it.
\cite{Eldridge2008} assumed that a flat distribution over $1< \log a_0/R_\odot<4$ \citep[also see][]{Tout2014},
where $a_0$ is the initial binary separation. In this work, we adopt the above distribution.

\subsection{Star Formation Rate}
\begin{figure}
\includegraphics[totalheight=3.0in,width=2.5in,angle=-90]{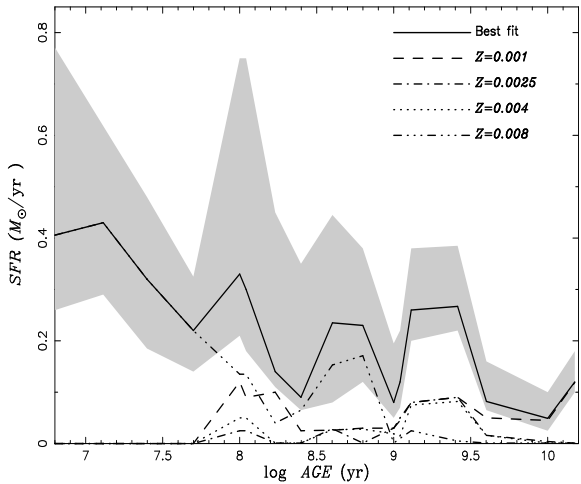}
\caption{Star formation history in
the LMC with the age from
\citet{Harris2009}. The best-fit SFR as a function of age
is showed with a solid line, and the uncertainty on the fit
is showed as an oblique-line area. Dashed, dash-doted, doted, dash-dot-dot-doted lines
present the contributions of $Z=0.001$, 0.0025, 0.004 and 0.008, respectively.
 }
\label{fig:sfr}
\end{figure}

The dust quantities in the LMC depend on the star formation rate (SFR).
According to the Magellanic Clouds Photometric Survey including 20 million LMC stars
and using the StarFISH analysis software,
\cite{Harris2009} reconstructed the SFR of the LMC during different ages.
\cite{Harris2009} found that the optimal solution includes the four metallicities
$Z=0.001, 0.0025, 0.004, 0.008$. The separate contributions of the four metallicities
during different ages are shown in Figure \ref{fig:sfr}.
Since we are interested in the dust quantities of the whole LMC, we average over
SFRs of individual regions for each age bin.

\subsection{Initial Binary Fraction in the LMC}
In our work, dust formation involves binary systems. Therefore, the binary fraction in all stars of the
LMC is very important. There is no observational evidence to help us in adopting a binary fraction.
\cite{Harris2009} simply assumed that the binary fraction in the LMC is 50\%. However,
it does not give solid information for the initial binary fraction. According to the theoretical model
of binary evolution, binary systems can become single stars if they undergo binary merger or they are
disrupted by supernovae. For example, for binary systems whose initial conditions are given in \S \ref{sec:monte},
about 38\% of binaries do not undergo Roche lobe mass transfer, and their evolutions are similar to
single stars. About 62\% of binaries undergo Roche lobe overflow at lease once,
in which about 73\% undergo CE evolution and can eject the whole CE at lease once,
while about 27\%  can not eject the CE and usually merge into single stars. In short, about 54\% of
these binary systems may be observed as single stars.

In this work, in order to check the effects of initial binary fraction on the dust budget in the LMC,
we use two different initial binary fractions: one is an initial binary fraction of
50\% (a half of SFR in the LMC is for binary systems and another half is for single stars), the other
is that the initial binary fraction is 100\%, that is, all stars are born in binary systems.

\subsection{Calculation of Dust Production Rates}
Using IMF, SFR and dust yields discussed in the previous sections,
we can theoretically calculate DPRs from the above three sources in the LMC.

For a single star, following
\cite{Zhukovska2013} and \cite{Schneider2014}, the DPRs of the type $j$ dust grains from AGB stars at time $t$ can be estimated by
\begin{equation}
\dot{M}_{\rm d}^{j}=\int^{m_{\rm U}}_{m_{\rm L}}\phi(m)\frac{SFR(t-\tau(m,Z))}{2}m^{j}_{\rm d}(m,Z){\rm d}m,
\label{eq:sd}
\end{equation}
where $\phi(m)$ is IMF, $SFR(t-\tau(m,Z))$ is SFR, $\tau(m,Z)$ is the lifetime of a star
with mass $m$ and metallicity $Z$ when it leaves AGB stage, and $m^{j}_{\rm d}(m,Z)$ is the mass of dust.
The lower integration limit, $m_{\rm L}$, is determined by a stellar mass with the lifetime which satisfies $\tau(m_{\rm L}) = t$,
and the upper mass-limit $m_{\rm U}= 8M_\odot$ is determined by stars becoming AGB stars.
The $m^{j}_{\rm d}(m,Z)$ can be obtained by the linear interpolation in
a given grid of dust yields from \cite{Zhukovska2008} or \cite{Schneider2014}.
Similarly, the DPRs of the type $j$ dust grains from SNe II at time $t$
can also be calculated by Eq.(\ref{eq:sd}), but the upper mass-limit $m_{\rm U}= 40M_\odot$ and
the $m^{j}_{\rm d}(m,Z)$ can be obtained from \cite{Todini2001}.

For a binary system in which the initial masses of primary and secondary are $m_1$ and $m_2$ and the initial binary
separation is $a_0$,
the DPR of the type $j$ dust grains from CE ejecta at time $t$ can be estimated by
\begin{equation}
\begin{array}{ll}
\dot{M}_{\rm d}^{j}=&\int^{m_{\rm U}}_{m_{\rm L}}\phi(m_1)\int^{m_{\rm 1}}_{m_{\rm L}}\phi(m_2)\int^{10}_{10^4}\varphi(a_0)\\
 &\frac{SFR(t-\tau(m_1,m_2,a_0))}{2}m^{j}_{\rm d}(m_1,m_2,a_0){\rm d}m_1{\rm d}m_2{\rm d}a_0,\\
\end{array}
\label{eq:bd}
\end{equation}
where $\tau(m_1,m_2,a_0)$ is the time of a binary system occurring CE evolution, and $m^{j}_{\rm d}(m_1,m_2,a_0)$ is
dust mass produced by CE ejecta. In order to calculate $m^{j}_{\rm d}(m_1,m_2,a_0)$, for the four metallicities ($Z=0.001,0.0025,0.004,0.008$), we calculate dust yields which are for different stellar masses ($m=1.0,1.5,2.0,2.5,3.0,3.5,4.0,4.5,5.0,5.5,6.0,8.0 M_\odot$) in different evolutional phases(HZ, the beginning of FGB, the end of FGB, the beginning of AGB and TPAGB). We do not consider massive stars due to mass function.
When a binary system undergoes CE evolution, by the linear interpolation of the donor' mass and age in the dust yields, we obtain $m^{j}_{\rm d}(m_1,m_2,a_0)$.

\subsection {Dust Production Rates}
\begin{figure*}
\begin{tabular}{cc}
\includegraphics[totalheight=3.2in,width=2.5in,angle=-90]{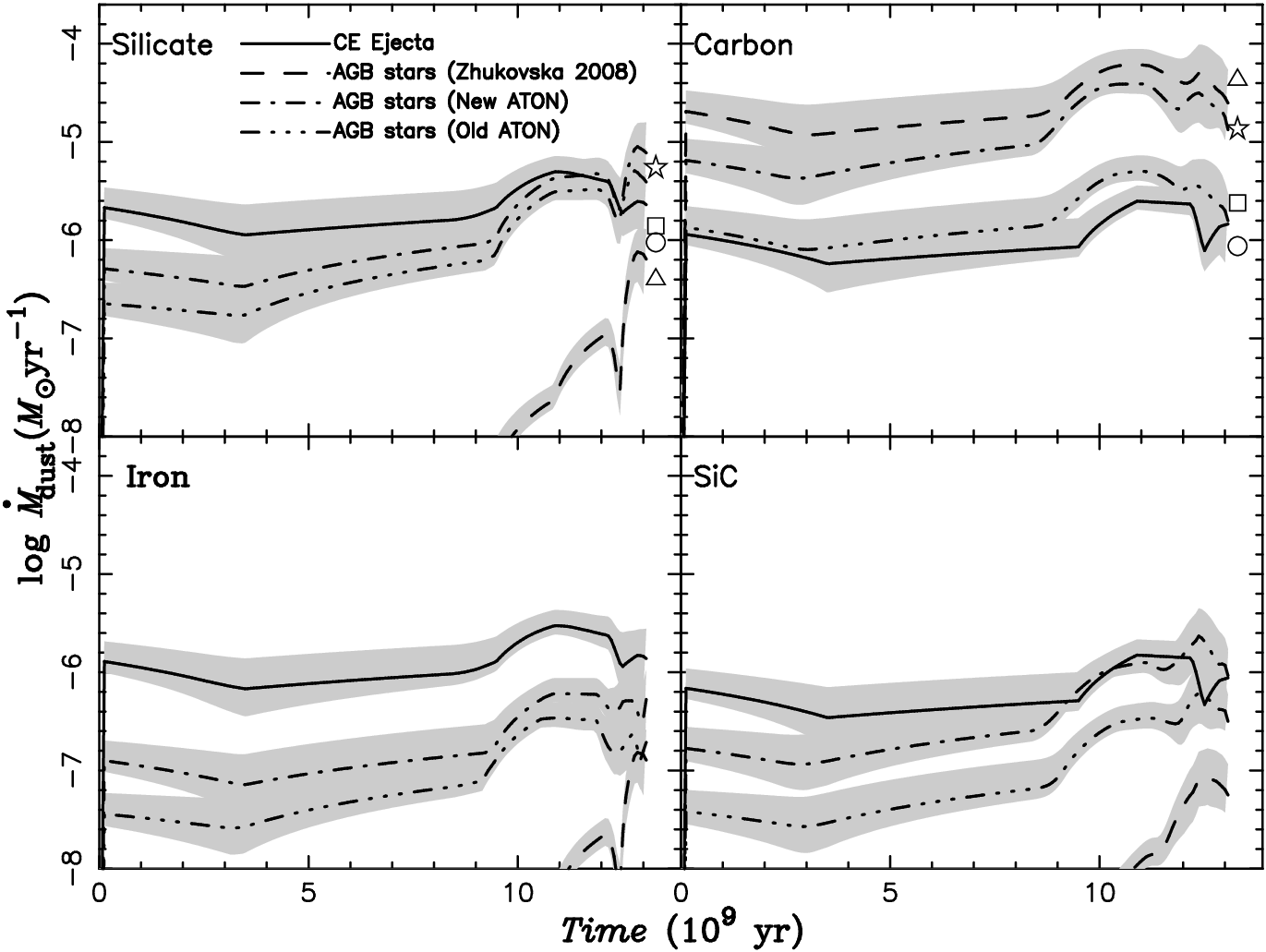}
&\includegraphics[totalheight=3.2in,width=2.5in,angle=-90]{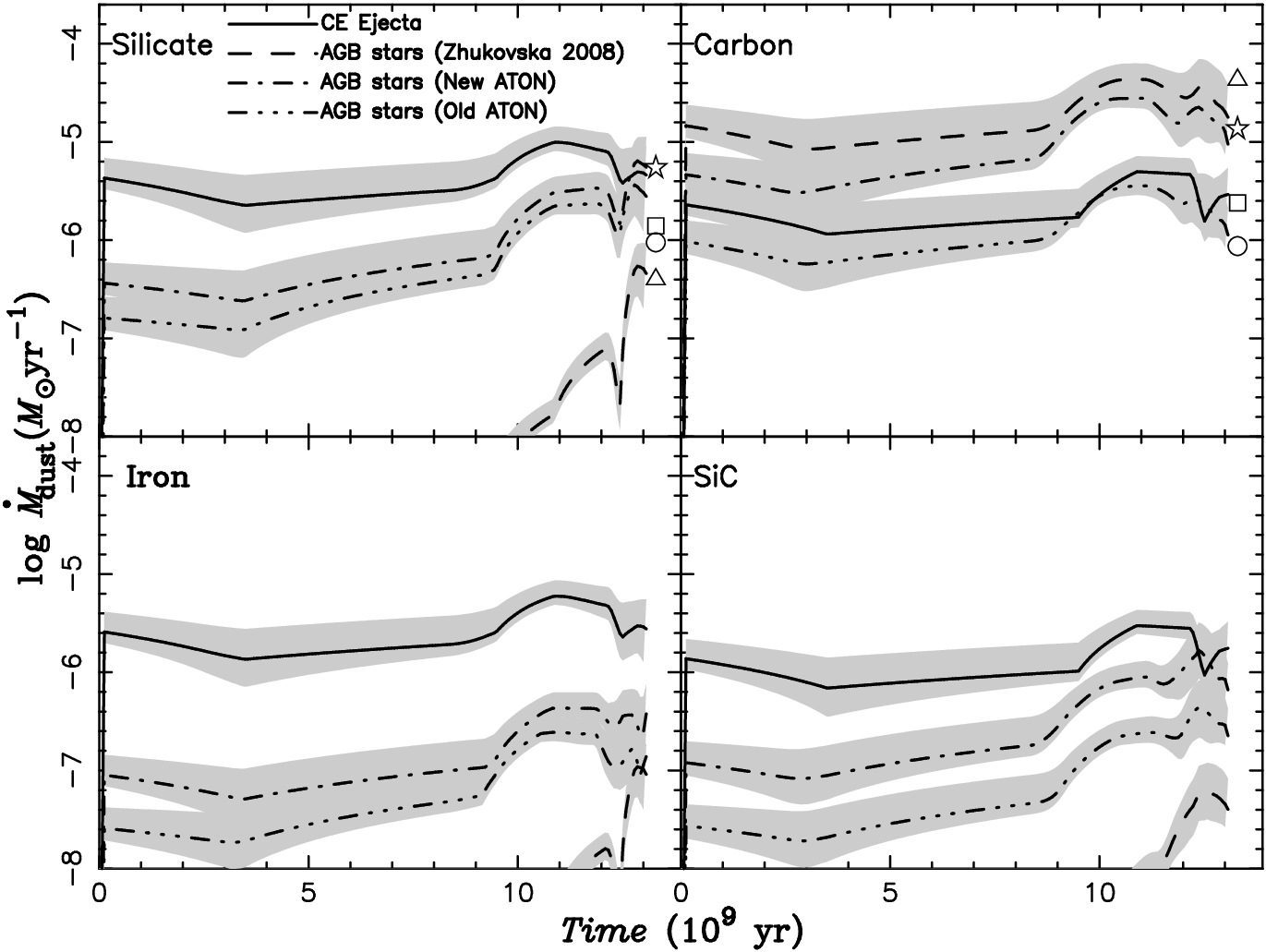}\\
\end{tabular}
\caption{The dust production rates (DPRs) of different dust species from CE ejecta and AGB stars in the LMC as
a function of time for metallicity-dependent star formation histories given by \citet{Harris2009}.
Left and right panels show that the initial binary fractions are 50\% and 100\%, respectively.
The solid, dashed, dash-doted and dot-dot-dot-dashed lines show the
predicted DPRs from CE ejecta, AGB stars using the dust yields of \citet{Zhukovska2008},  new and old ATON, respectively.
The shaded regions around lines represent the uncertainty on the best-fit star formation
histories. The DPRs of O-rich and C-rich AGB stars observationally estimated by \citet{Matsuura2009}, \citet{Srinivasan2009}, \citet{Boyer2012} and
\citet{Riebel2012} are represented by triples, stars, squares, and circles, respectively.}
\label{fig:dpr}
\end{figure*}

Based on observational data of a number of surveys, DPRs from C-rich and O-rich AGB stars in the LMC had been
estimated by \cite{Matsuura2009}, \cite{Srinivasan2009}, \cite{Boyer2012} and \cite{Riebel2012}.
Unfortunately, as Figure \ref{fig:dpr} shows, there is a large scatter between the above observationally estimated DPRs.
DPRs from C-rich AGB stars estimated by \cite{Matsuura2009} are larger than
$4.3\times10^{-5} M_\odot$ yr$^{-1}$ (up to $1.0\times10^{-4} M_\odot$ yr$^{-1}$), while they
 are $2.4\times10^{-6} M_\odot$ yr$^{-1}$ in \cite{Srinivasan2009}, $8.7\times10^{-7} M_\odot$ yr$^{-1}$ in \cite{Boyer2012}, 5.2---5.7$\times10^{-6} M_\odot$ yr$^{-1}$ in \cite{Riebel2012}, respectively.
Similar scatter appears in DPRs from O-rich AGB stars.
Differences of the DPRs observationally estimated mainly result from various approaches and source classifications (Details can be seen in
\cite{Zhukovska2013}).

Similarly, there is also a large difference between the theoretically estimated DPRs.
Using the dust yields of  AGB stars in \cite{Zhukovska2008}, \cite{Zhukovska2013} predicted that the DPRs from C-rich and O-rich AGB stars
in the LMC
are 5.7$\times10^{-5}M_\odot$ yr$^{-1}$ and 1.3$\times10^{-6}M_\odot$ yr$^{-1}$, respectively. However,
using the dust yields of  AGB stars from old ATON, \cite{Zhukovska2013} calculated
that they were 6.3$\times10^{-6}M_\odot$ yr$^{-1}$ and 1.1$\times10^{-5}M_\odot$ yr$^{-1}$, respectively. The uncertainty introduced
by different models for AGB stars is up to about 10. Using the dust yields of  AGB stars from \cite{Zhukovska2008}, old and new ATON,
\cite{Schneider2014} also predicted the DPRs in the LMC. They found that the DPR from C-rich AGB stars calculated by the dust yields of  AGB stars from \cite{Zhukovska2008} is
close to that calculated by new ATON, while the DPR from O-rich AGB stars by new ATON is close to that by old ATON.
The main reason behind these differences is in the treatment of convective borders and in the efficiency of the convective models used, that affect the extent of the third dredge-up and the strength of hot bottom burning (Details can be seen in
\cite{Schneider2014}).
However, to our knowledge, all theoretical estimates did not consider the effect of binary interaction on the DPRs in the LMC.
In this work we compare the DPRs of AGB stars in the LMC with those from \cite{Zhukovska2008}, old and new ATON.
Simultaneously, we also estimates the DPRs of CE ejecta in the LMC.
As Figure \ref{fig:dpr} shows, the initial binary fraction changing from 50\% to 100\% introduces an uncertainty of a factor of
about 1.5 --- 2 to the DPRs from AGB stars and from CE ejecta, respectively. These uncertainties are much smaller than these
differences between different observational or theoretical estimates.

Due to large scatters between observationally estimated or theoretically predicted DPRs from AGB stars,
we can not determine whether the DPRs from CE ejecta are significant. However, we find that
CE ejecta can efficiently produce silicate and iron grains, while carbon grains are negligible.
The main reason is that in our models most of CE events occur before the third dredge-up works efficiently.

\begin{figure*}
\begin{tabular}{cc}
\includegraphics[totalheight=3.2in,width=2.5in,angle=-90]{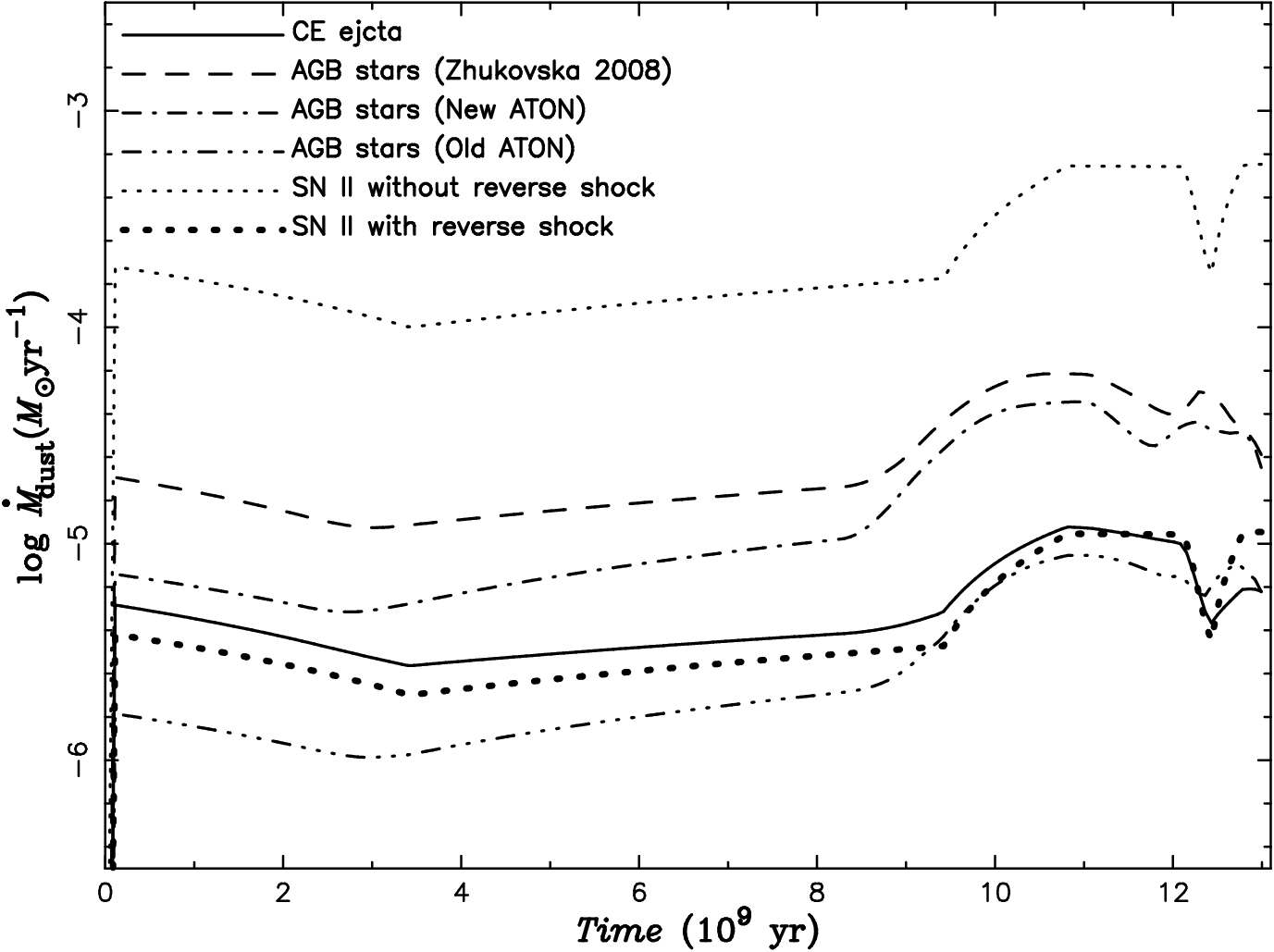}
&\includegraphics[totalheight=3.2in,width=2.5in,angle=-90]{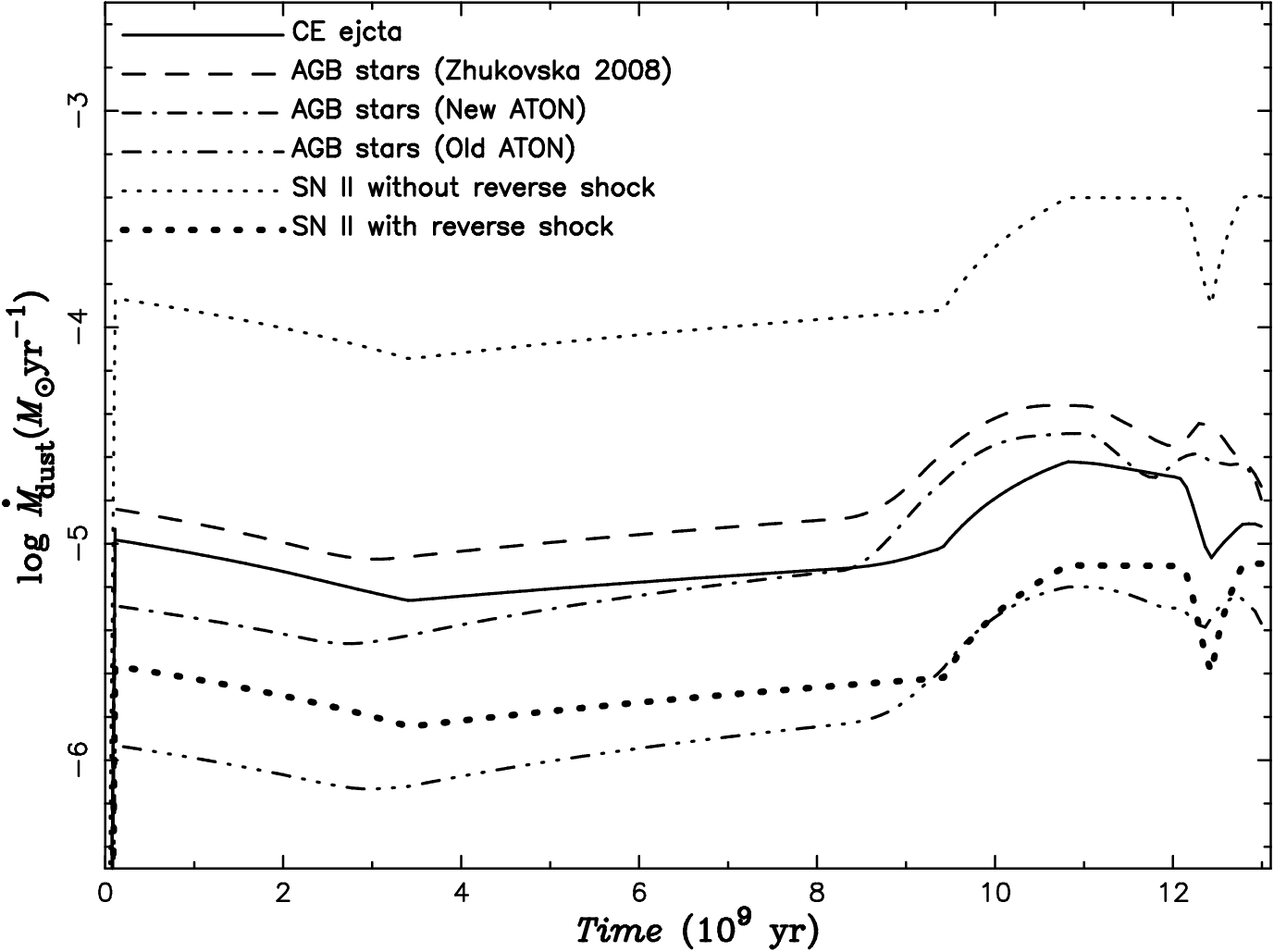}\\
\end{tabular}
\caption{The total DPRs from CE ejecta, AGB stars and SNe II with the
age of the LMC by using the best-fit star formation
histories. Left and right panels show that the initial binary fractions are 50\% and 100\%, respectively.
The solid, dashed, dash-doted and dot-dot-dot-dashed lines show the
predicted DPRs from CE ejecta, AGB stars using the dust yields of \citet{Zhukovska2008} and the new ATON in \citet{Schneider2014}, respectively.
Thin doted lines represent the DPRs from SNe II using the dust yields of \citet{Todini2001}, while
thick doted lines use 2\% of the dust yields of \citet{Todini2001}.  }
\label{fig:dprsn}
\end{figure*}

In order to compare the contributions of different sources to the DPRs in the LMC,
Figure \ref{fig:dprsn} shows the total DPRs from CE ejecta, AGB star and SNe II with the
age of the LMC by using the best-fit star formation histories. If we do not
consider the destruction by reverse shock, the majority of dust grains in the LMC originate from
SNe II. However, if only 2\% of dust grains can survive after reverse shock,
compared with AGB stars calculated by \cite{Zhukovska2008} and new ATON,
the contribution of SNe II to the total DPRs is negligible. In the simulation with an initial binary
fraction of 50\%, the DPR from CE ejecta is close to that calculated by old ATON,
but it is only 1/4 of those calculated by \cite{Zhukovska2008} and new ATON.
However, it is about 4 times of the former and about 3/4 times of the latter in the
simulation with an initial binary fraction of 100\%. One should take note that the assumption of 100\% binary fraction
is realistic. Here, we only want to estimate an upper limit of the DPR from CE ejecta.
More importantly, most of dust grains calculated by \cite{Zhukovska2008} and new ATON  are
carbon. Formation of C-rich AGB stars deeply depend on a complex physical process:
the third dredge-up. In new ATON model and \cite{Zhukovska2008},
the third dredge-up is more efficient, and C-rich AGB stars form more easily. However,
in old ATON, it is difficult for a star to evolve into C-rich AGB star, and the DPR from C-rich AGB stars
in the LMC is negligible\citep{Zhukovska2008,Schneider2014}.

In short, although the DPRs in the LMC estimated by observational data or theoretical models are very uncertain,
we consider that the contributions of CE ejecta must be included when the total DPRs in the LMC are calculated.



\subsection{Total Dust Budget}
\begin{figure*}
\begin{tabular}{cc}
\includegraphics[totalheight=3.2in,width=2.5in,angle=-90]{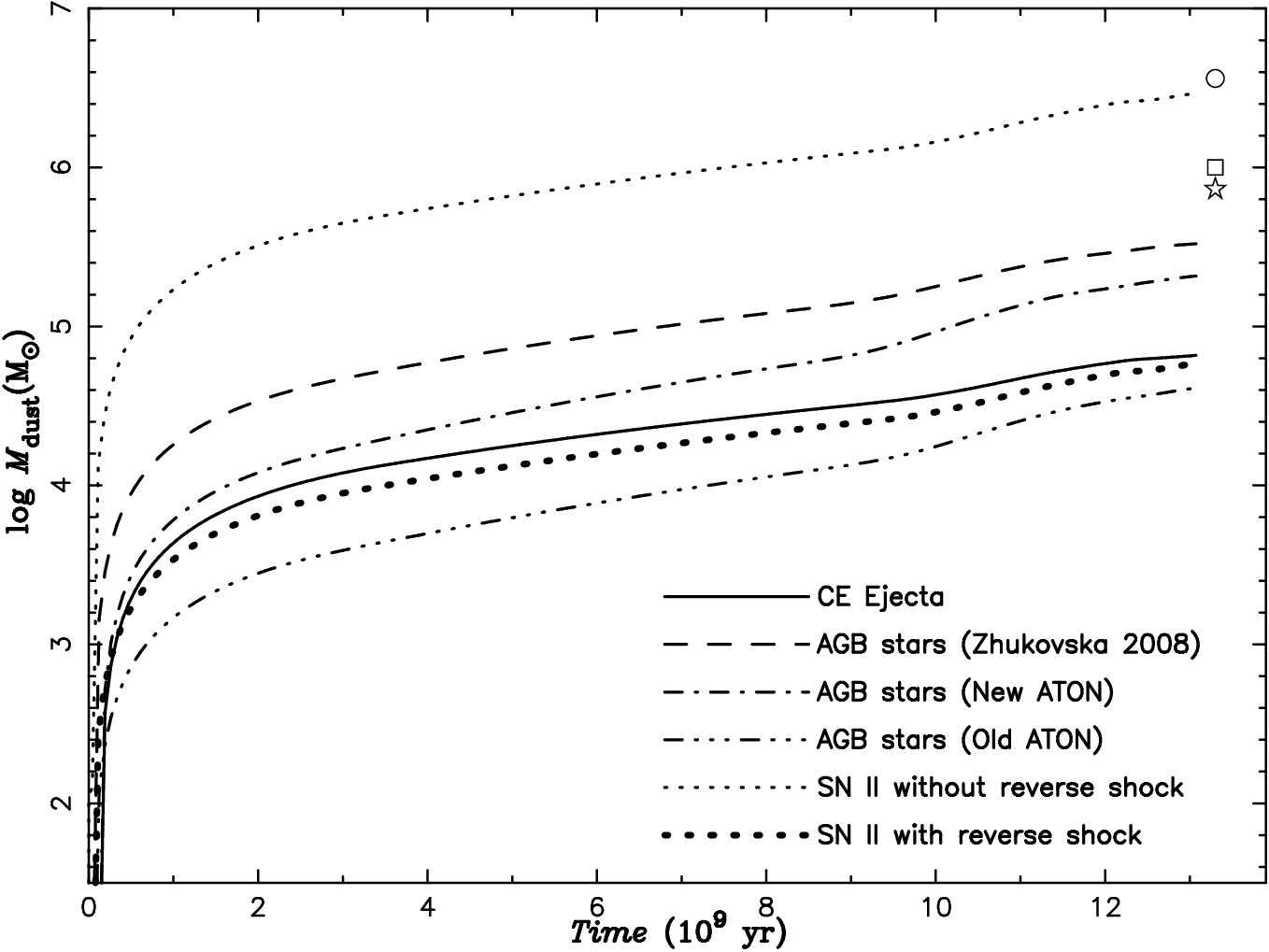}
&\includegraphics[totalheight=3.2in,width=2.5in,angle=-90]{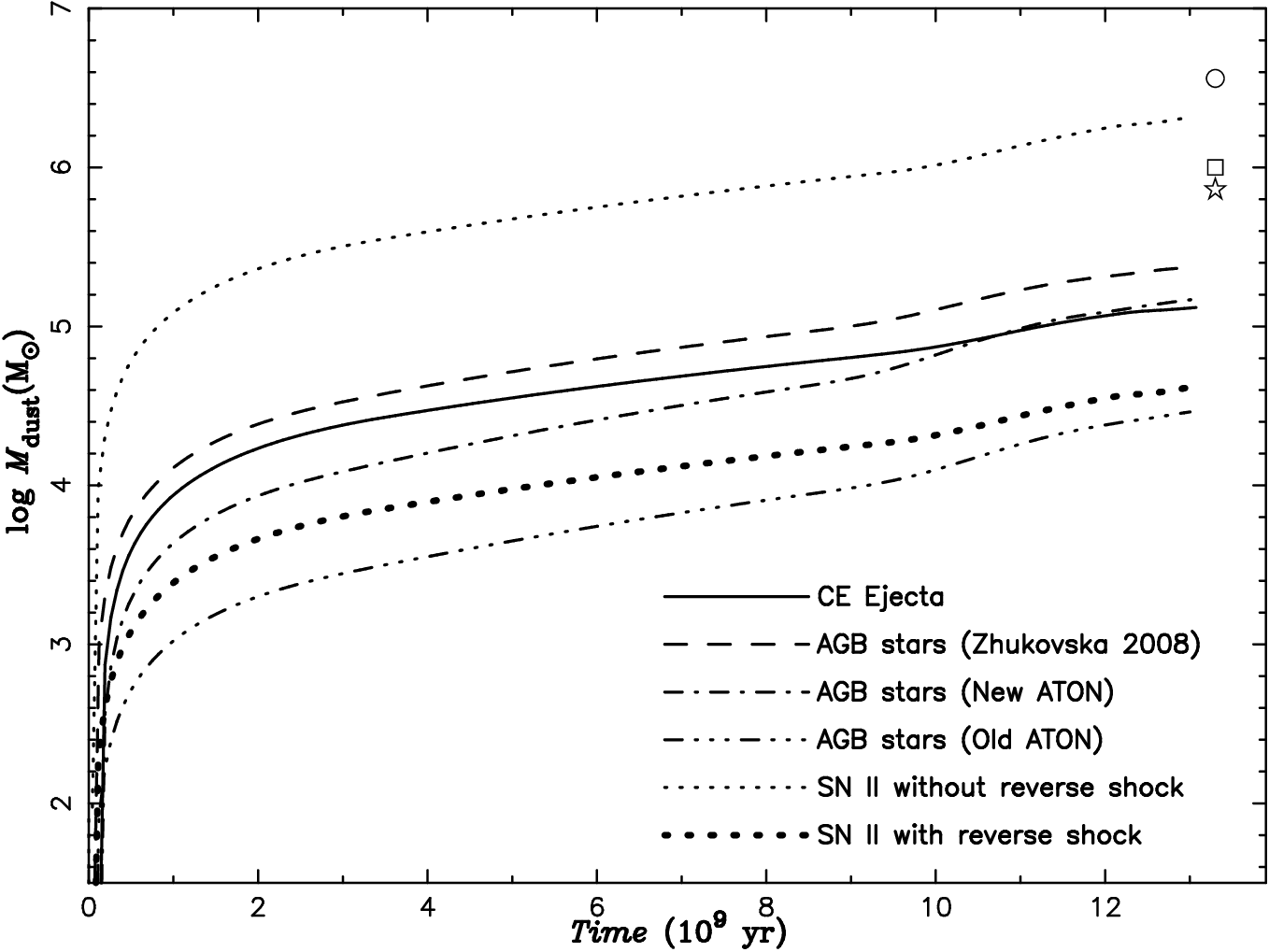}\\
\end{tabular}
\caption{Similar with Figure \ref{fig:dprsn}, but for time evolution of total dust masses
from CE ejecta, AGB stars and SNe II. The observational estimates presented by circle, square and star are taken from
\citet{Bot2010}, \citet{Skibba2012} and \citet{Gordon2014}, respectively.}
\label{fig:dustsn}
\end{figure*}

The total dust mass, $M_{\rm dust}$, is very important for understanding the dust formation, evolution and destruction in the LMC.
Therefore, there are many literatures to estimate $M_{\rm dust}$ in the LMC by using some observational
data\citep[e. g.,][]{Bot2010,Skibba2012,Gordon2014}. However, the total dust masses estimated are different:  Using
sub-millimeter excess of the LMC and fitting the dust models of \citet{Draine2007}, \citet{Bot2010} obtained $M_{\rm dust}$ of
$3.6\times10^6M_\odot$; According to {\it Herschel} HERITAGE data, \citet{Skibba2012} predicted
$M_{\rm dust}\sim 1.1\times10^6M_\odot$; Based on the HERITAGE {\it Herschel} Key Project photometric data,
\cite{Gordon2014} obtained $M_{\rm dust}$ of $7.3\times10^5M_\odot$, which is a factor of about 5 lower than previous results.
These differences maybe come from different dust models, the fitting techniques, the wavelength range, and the factor of environments \citep{Gordon2014}.
As Figure \ref{fig:dustsn} shows, the total dust mass produced by AGB stars or CE ejecta is much lower
than the above observational estimates. Of course, we do not know whether dust produced by SNe II is
dominant because theoretical models and observational estimates can hardly give any certain value for the dust yields of SNe II.

Furthermore, dust grains can be destroyed by the blast wave produced by SNe. Therefore,
their existence should have time scale. According to the properties of the
dust material and the occurrence rate of SNe in the LMC, \cite{Zhukovska2013} estimated the time scale of dust destruction.
They found that the time scale for silicate grains is about 0.8 Gyr, and it is about 1.1 Gyr for carbonaceous grains.
This means that about 10\% of dust masses showed in Figure \ref{fig:dustsn} can survive if we consider the dust destruction.
It is possible that AGB stars and CE ejecta are not main dust sources and dust can grow in the ISM of the LMC \citep{Draine2009}.

\begin{figure*}
\begin{tabular}{cc}
\includegraphics[totalheight=3.2in,width=2.5in,angle=-90]{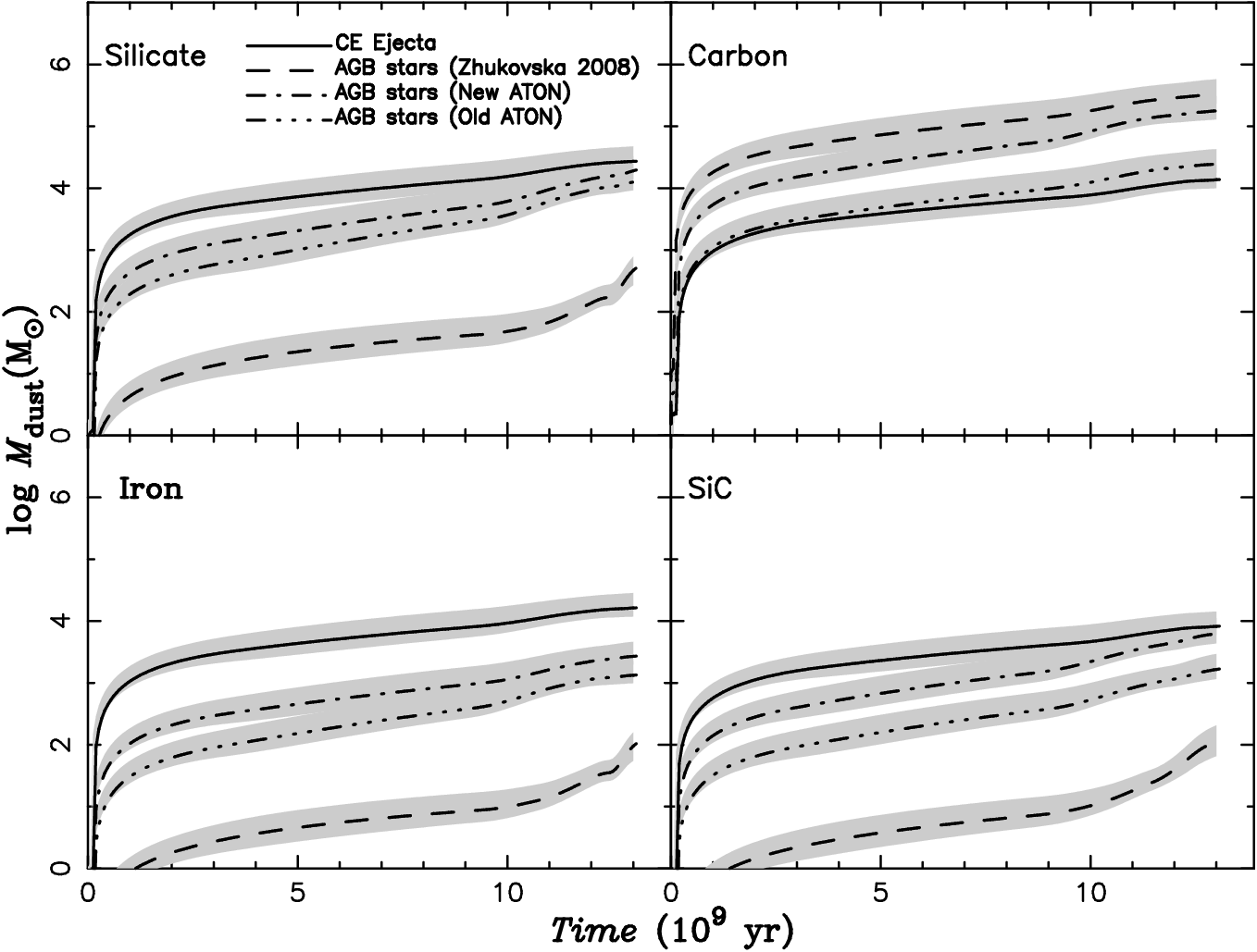}
&\includegraphics[totalheight=3.2in,width=2.5in,angle=-90]{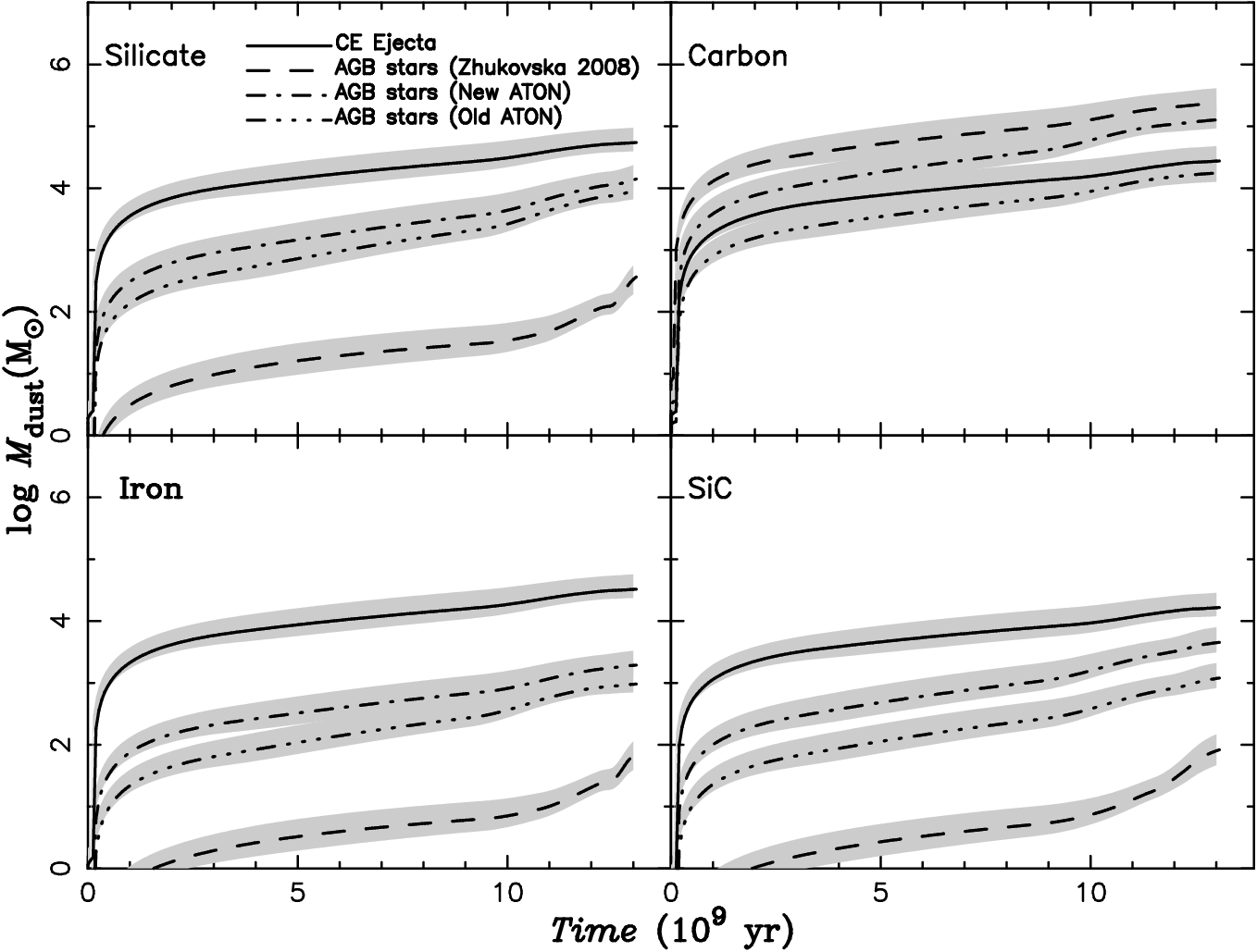}\\
\end{tabular}
\caption{Similar with Figure \ref{fig:dpr}, but for time evolution of accumulated masses
of different dust species from CE ejecta and AGB stars.}
\label{fig:dustq}
\end{figure*}

 As discussed in Introduction, according to the estimates of \cite{Weingartner2001}, we indirectly guess that
the mass ratio of carbon to silicate grains in the LMC is about 1:4. However, to our knowledge,
there is no direct observations to give the compositions of different dust species in the LMC.
It is very important for us to understand the dust origin. We theoretically predict the time evolution of accumulated masses
of different dust species from CE ejecta and AGB stars in Figure \ref{fig:dustq}.  Obviously,
in the model of \cite{Zhukovska2013}, most of dust grains are carbon and others are negligible.
In the model of new ATON, the mass fractions of carbon and silicate grains are about 82\% and 13\%, respectively, and the left are
mainly SiC grains. In the  model of old ATON, the mass fractions of carbon and silicate grains are about 55\% and 35\%, respectively.
The remaining are SiC and iron grains. However, in CE ejecta, about 50\% of dust grains are silicate, 20\% of them are iron
 and the mass fractions of carbon and SiC grains are about 17\% and 13\%, respectively. Not like AGB stars which mainly form C-rich grains,
 CE ejecta chiefly produces O-rich dust grains.

 Therefore, with the further observations, we believe that the compositions of different dust species in the LMC will
 be found soon. Then, we can estimate which between AGB stars and CE ejecta is more important dust source. Of course,
 we also should consider the effect of SNe II and dust growth in the ISM.

\section{Conclusions}
In this work, we calculate the contributions of dust produced by
AGB stars, CE ejecta and SNe II to the total dust budget in the LMC.
Because different observations still have large scatter and theoretical models
greatly depend on complex physical processes, we can not determine which is the main dust
source in the LMC. However, more than 50\% of stars in the LMC are in binary systems. Many of them
may have or will undergo CE evolutions. According to our model, CE ejecta can efficiently produce dust.
Therefore, we must consider the contributions of dust produced by CE ejecta on estimating the total
dust budget.

In our simulations, the DPRs of AGB stars in the LMC are between about $2.5\times10^{-5}M_\odot{\rm yr^{-1}}$
and $4.0\times10^{-6}M_\odot{\rm yr^{-1}}$. The uncertainty mainly results from different models for the dust yields of AGB stars.
These results are within the large scatter of several observational estimates.
The DPRs of CE ejecta are about $6.3\times10^{-6}M_\odot{\rm yr^{-1}}$ (The initial binary fraction is 50\%).
Compared the theoretically predicted DPRs of
AGB stars,  the DPRs of CE ejecta are significant or dominated. The contribution of SNe II is
very uncertain. Compared with SNe II without reverse shock, the DPRs of AGB stars and CE ejecta are negligible.
However, if only 2 \% of dust grains produced by SNe II can survive after reverse shock, the contribution of SNe II
is very small.

The total dust mass of the LMC estimated by observations is between $7.3\times10^5M_\odot$ and $3.6\times10^6M_\odot$.
In our work, the total dust mass produced by AGB stars in different models are between $2.8\times10^4M_\odot$ and $3.2\times10^5M_\odot$,
and those produced by CE ejecta are about $6.3\times10^4M_\odot$.
In our calculations, the total dust masses produced by SNe II are vary uncertain, and they change from $2.8\times10^3M_\odot$ to $2.8\times10^6M_\odot$.
If the destruction of blast wave from SNe is considered, compared with the observational estimates,
the total dust masses produced by AGB stars, CE ejecta and SNe II are negligible.  Therefore,
on estimating the total dust mass, the growth of dust in the ISM must be considered.

Not only observational estimates but also theoretical predictions found that AGB stars in the LMC mainly produce
carbon grains. Based on our simulations, most of dust grains forming in CE ejecta are silicate and iron grains.
Therefore, with the further observations, the compositions of different dust species will be found, which
may be very important for understanding the dust origins and  the contributions of dust produced by AGB stars and CE ejecta.

\section*{Acknowledgments}
We thank an anonymous referee for his/her comments which helped to
improve the paper.
This work was supported by
XinJiang Science Fund for Distinguished Young Scholars under Nos. 2014721015 and 2013721014,
the National Natural Science Foundation
of China under Nos. 11473024, 11363005 and 11163005.


\bibliographystyle{mn2e}
\bibliography{lglmn}
\bsp

\label{lastpage}

\end{document}